\documentclass[12pt]{article}

\advance\textwidth by 0.5in
\advance\hoffset by -0.25in

\usepackage{epsf}

\def\lsim{ \,\, \vcenter{\hbox{$\buildrel{\displaystyle <}\over\sim$}}
 \,\,}
\newcommand{\be}{\begin{eqnarray}}
\newcommand{\ee}{\end{eqnarray}}
\newcommand{\non}{\nonumber\\}
\newcommand{\inline}[1]{\noalign{\hbox{#1}}}
 \newcommand{\absol}[1]{\left| #1 \right|}
 \newcommand{\GeV}{\hbox{\,GeV}}
 \newcommand{\MeV}{\hbox{\,MeV}}
 \newcommand{\fm}{\hbox{\,fm}}
 \newcommand{\Tr}{\hbox{Tr\,}}
 \newcommand{\eqn}[1]{Eq.~(\ref{#1})}

\begin{document}
\title {\bf Production of $\eta'$ from Thermal Gluon Fusion}
\author
{
 Sangyong Jeon
 \\
 {\small\it Department of Physics}\\
 {\small\it McGill University}\\
 {\small\it 3600 University Street, Montreal, QC H3A-2T8, Canada}\\
 {\small\it and}\\
 {\small\it RIKEN-BNL Research Center}\\
 {\small\it Brookhaven National Laboratory}\\
 {\small\it Upton, NY 11973}\\
}

 \maketitle

 \begin{abstract}
 We study the production of $\eta'$ from 
 hadronizing thermal gluons using
 recently proposed $\eta'-g-g$ effective vertex. 
 The $\eta'$ yield is found to be sensitive to the initial
 condition.  At RHIC and LHC, the enhancement is 
 large enough to be easily detected.
 \end{abstract}

 \newpage

 \section{Introduction}

 If the hadronic Lagrangian is symmetric under flavor $U(3)$ which is
 spontaneously broken, then we would have 9 pseudoscalar Goldstone bosons.
 In reality, we have 8 light mesons $(\pi, K, \eta)$ 
 corresponding to the octet of
 $SU(3)$ and one heavy meson, $\eta'$.
 
 The pseudoscalar flavor singlet $\eta'$ is a remarkable resonance.
 Its large mass poses the $U_A(1)$ problem and 
 its possible resolution relates its mass to the
 topological charge of the QCD vacuum 
 and to the properties of the instanton liquid  
 \cite{Belavin:1975fg,Witten:1979vv,Veneziano:1979ec,'tHooft:1976up}.
 
 In heavy ion physics, the $\eta'$ meson
 is a good probe because it has a lifetime
 $(1000\fm)$ long
 compared to the typical lifetime of a fireball produced by a
 collision of relativistic heavy ions.  This was exploited by 
 the authors of Ref.~\cite{Kapusta:1996ww} 
 who studied possible lowering of the $\eta'$ mass by
 the disappearance of the instanton liquid at high temperatures.
 In it, the authors argued that even in dense matter the $\eta'$ 
 meson may decouple from the rest of the matter.

 Recently, there was a surge of interest in $\eta'$ 
 in the study of $B$-meson decays and search
 for new physics 
 \cite{Kagan:1997qn,Atwood:1997bn,Muta:2000tc,Ali:1998ex,Hou:1998wy}.
 In some of these studies, the axial anomaly relation 
 \be
 \partial^\mu J^0_{5\mu}
 =
 2 N_f {g^2\over 16\pi^2} \, 
 {\rm Tr}\,\left( G_{\mu\nu} \tilde{G}^{\mu\nu}\right)
 \ee
 is interpreted to imply that the gluons and $\eta'$ have an effective
 Wess-Zumino-Witten type 
 interaction vertex~\cite{Atwood:1997bn,Hou:1998wy}
 (See also \cite{Damgaard:1994sx})
 \be
 M_{\lambda\gamma} \delta^{ab}
 = H(p^2, q^2, P^2)\,
 \delta^{ab} \,
 \epsilon_{\mu\nu\alpha\beta}\, 
 p^\mu \,
 q^\nu \,
 (\epsilon_p^\alpha)_\lambda \,
 (\epsilon_q^\beta )_\gamma
 \label{eq:vertex}
 \ee
 where $p, q$ are the gluon momenta and
 $(\epsilon_{p,q})_{\lambda,\gamma}$ are the
 corresponding gluon polarization vectors and the superscripts $ab$
 denote the color indices of the two gluons.
 The momentum of $\eta'$ is denoted by $P$ throughout the paper.
 By studying $J/\psi \to \eta'\,\gamma$ decay process,
 Atwood and Soni~\cite{Atwood:1997bn} found that this process is
 dominated by on-shell gluons and obtained
 \be
 H_0 \equiv H(0,0,M_{\eta'}^2) \approx 1.8\,\hbox{GeV}^{-1}
 \ee

 The above $gg\eta'$ effective 
 vertex is interesting in many ways.  First,  
 since the $\eta'$ mass is almost $1\GeV$, at least one of
 the gluon momenta involved in the vertex should be greater than 
 $0.5\,\GeV$.  Therefore the gluon momenta are not soft compared to
 the temperatures achievable in heavy ion collisions.
 Second, this is a rare occasion when we know (at
 least we can parameterize) how to fuse two on-shell gluons and form a hadron.
 There are models in the literature that relates {\em constituent} quarks to
 the hadrons, but to the author's knowledge, there is no other known
 matrix element between gluons and a known hadron state. 

 In this paper, we exploit these unique circumstances and study the
 production of the $\eta'$ mesons from a hadronizing quark-gluon plasma. 
 One question we have to answer before we proceed is how the interaction
 strength $H_0 = H(0,0,M_{\eta'}^2)$ changes as the temperature
 increases.  In the case of anomalous coupling of photons to $\pi^0$,
 it is known that the coupling strength vanishes in the chiral limit 
 \cite{Pisarski:1997bq,Kumar:2000xm}
 although the axial anomaly itself is not affected by the temperature
 \cite{Itoyama:1983up,Schafer:1996hv,Elmfors:1997fx}.
 As the chiral symmetry restoration temperature is approximately the same 
 as the deconfinement temperature, one should then ask if the $gg\eta'$
 strength also vanishes as the temperature rises above the critical
 temperature.
 
 A partial answer to this question may be given by the result of
 Ref.~\cite{Kumar:2000xm}. 
 In Ref.~\cite{Kumar:2000xm}, 
 the authors carefully analyzed the triangle diagram
 contribution of $\pi^0 \to \gamma\gamma$ decay at finite temperature
 and obtained
 \be
 g_{\pi\gamma\gamma} =  {m_q\over T^2} e^2 g_{\pi q\bar q} 
 \,{\cal F}\left[
 g_{\pi q\bar q}, \alpha, \alpha\,\ln 1/e, ...
 \right]
 \ee
 where $m_q$ is the constituent quark mass,
 $e$ is the electromagnetic coupling constant and
 $g_{\pi q\bar q}$ is the $\pi^0$ quark anti-quark coupling constant.
 ${\cal F}$ is a finite function of coupling constants.
 If the QCD anomalous coupling of the gluons to
 $\eta'$ is similar to $\pi^0\gamma\gamma$ coupling, 
 the above expression can be rewritten as 
 \be
 g_{\eta'gg} = H_0 \sim  {m_q\over T^2} 
 \label{eq:rough_est}
 \ee
 since the coupling constant involved are all $O(1)$.
 In a quark-gluon plasma,
 the $u$ and $d$ quark masses vanish.  However, the
 strange quark mass does not vanish.  Since 
 $\eta' \approx (u\bar u + d\bar d + s\bar s)/\sqrt{3}$, this indicates
 that the $\eta'gg$ vertex does not necessarily vanish in this limit.  
 Rather, it will be proportional to the strange quark mass.
 Furthermore, the $\eta'$ mesons produced by the fusing gluons will be
 dominated by $s\bar s$ component.

 There is no doubt that for a more quantitative calculation, we need
 to calculate the $\eta'gg$ vertex at finite temperature
 with full finite temperature complications. 
 In this work, we will simply take the coupling 
 to be the same as the vacuum value, $H_{T} = 1.8\GeV^{-1}$  
 but take $\eta'$ from gluon fusion to be in a $s\bar s$ state so that 
 $M_{\eta'} \approx 0.7\GeV$ \cite{Kapusta:1996ww}.  
 If we follow the analysis in 
 Ref.~\cite{Kapusta:1996ww}, the mass of $\eta'$ may be as small as the
 pion mass at high temperature.
 But in this study, we take the above conservative estimate.  
 The finite temperature correction is currently under investigation.

 \section{Kinetic Theory Approach}

 Kinetic equations are an statement about the change of the phase space
 density in time
 \be
 {df\over dt}
 =
 (\hbox{Gain Rate})
 -
 (\hbox{Loss Rate})
 \ee
 To write down a Boltzmann equation for $\eta'$ distribution function,
 it is easiest to start with the decay rate.  In terms of the matrix
 element, 
 the decay rate of $\eta'$ to two gluons of the opposite colors and 
 different polarizations is given by 
 \be
 d\omega_{\eta'\to gg} 
 =
 \delta^{ab}
 {1\over 2}
 {1\over 2E_P}
 \absol{M_{\lambda\gamma}}^2\,
 {d^3p\over (2\pi)^3 2 p}\, 
 {d^3q\over (2\pi)^3 2 q}\, 
 (2\pi)^4 \delta^4(p + q - P)
 \ee
 where $p$ and $q$ are the gluon momenta and $P$ is the $\eta'$ momentum.
 The first factor of $1/2$ is the symmetry factor.
 Summing over all final states gives the total decay rate.
 It is then convenient to define
 \be
 \absol{M}^2_{\eta'\to gg}
 & = &
 \sum_{ab}\delta^{ab}
 \sum_{\lambda,\gamma} \absol{M_{\lambda\gamma}}^2
 \ee
 It is not hard to show 
 \be
 \absol{M}^2_{\eta'\to gg}
 =
 4\absol{H_0}^2\, M_{\eta'}^4
 \ee
 using the identities
 \be
 & 
 \displaystyle
 \sum_\lambda (\epsilon_p^{\alpha})_\lambda^* (\epsilon_p^\zeta)_\lambda
 =
 -g^{\alpha\zeta} + a p^\alpha p^\zeta
 \label{eq:eps_sum}
 &
 \\
 &
 \epsilon_{\alpha\beta\mu\nu} \,
 \epsilon^{\alpha\beta\rho\sigma} 
 =
 2
 \left(g^\mu_\sigma\, g^\nu_\rho
 -
 g^\mu_\rho\, g^\nu_\sigma
 \right)
 &
 \ee
 and the on-shell conditions
 $p^2 = q^2 = 0$ and $(p+q)^2 = M_{\eta'}^2$.
 The $ap^\alpha p^\zeta$ term in \eqn{eq:eps_sum}
 does not contribute to $M_{\lambda\gamma}$
 due to the anti-symmetric property of $\epsilon_{\alpha\beta\mu\nu}$.

\begin{figure}[t]
  \epsfxsize=10cm
  \centerline{\epsfbox{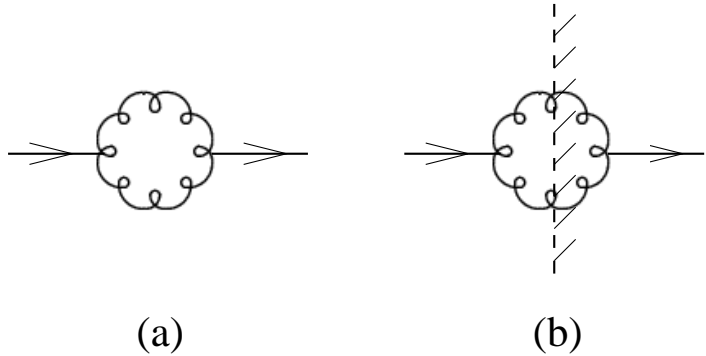}}
  \caption{
  Feynman graphs for the retarded self-energy of $\eta'$.  
  }
  \label{fig:self_energy}
 \end{figure}

 Employing the principle of detailed balance, we then write
 the Boltzmann equation for the phase space density of $\eta'$ as 
 \be
 \lefteqn{
 \partial_t f_{\eta'}(P) + {\bf v}\cdot\nabla f_{\eta'}(P)
 =
 {1\over 2}
 {1\over 2E_P}
 \int {d^3 p\over (2\pi)^3 2 p} {d^3 q\over (2\pi)^3 2 q}\,
 (2\pi)^4 \delta(p + q - P)
 } & & 
 \non
 & & \qquad {} 
 \times
 \absol{M}_{gg\to\eta'}^2\, 
 \left[
 f_g(p)\, f_g(q)\, (1 + f_{\eta'}(P)) 
 -
 (1+f_g(p))\, (1+f_g(q))\, f_{\eta'}(P) 
 \right]
 \label{eq:Boltzmann}
 \non
 \ee
 where ${\bf v} = {\bf P}/E_P$.
 Here it is understood that the distribution functions depend on
 the space time.  
 In the Boltzmann equation,
 the first term in the collision integral
 describes the production of
 $\eta'$ from the gluons and
 the second term describes the decay of $\eta'$
 into two gluons.  These collision terms are essentially the imaginary part
 of the retarded self-energy of $\eta'$ \cite{Jeon:1995if}
 depicted in Fig.~\ref{fig:self_energy} (b). 

 In a series of papers \cite{Jeon:1995if,Jeon:1996zm}, 
 it was shown that in a thermal
 medium, the real part of the self-energy must be also included
 in the mass parameter appearing in the Boltzmann equation. 
 With the effective vertex \eqn{eq:vertex},
 one can easily calculate the one-loop self-energy represented by the
 Feynman diagrams in Fig.\ref{fig:self_energy}.
 The details of their evaluation is presented in
 Appendix~\ref{app:self_energy}.
 In the present case, it turned out that the thermal correction is 
 negligibly
 small up to $T \approx 0.5\GeV$.  Therefore we can safely ignore it for
 our purposes.

 Before proceeding to analyze the Boltzmann equation, we must ask if we
 can use the Boltzmann equation in a quark gluon plasma.
 In other words, can $\eta'$ exist in a quark gluon plasma as a
 quasi-particle?
 It is possible that an excitation with the same quantum numbers as
 $\eta'$ can exist in the plasma (for instance see
 \cite{Kapusta:1996ww})
 but its width may be too broad to be a quasi particle.
 To be conservative, we apply the Boltzmann equation starting
 only from one relaxation time before the hadronization time 
 unless the hadronization time is shorter than the relaxation time.

 If the steady state is reached during the evolution,
 then the Boltzmann equation dictates
 that distribution functions become Bose-Einstein functions.  
 In this case, the distribution of $\eta'$ at the hadronization time
 will be simply the Bose-Einstein distribution with the temperature of
 $T_c \approx 0.17\GeV$.  This in itself is an interesting conclusion
 because there are surely other hadronization processes and hadronic
 processes that produce additional $\eta'$.  Since
 $\eta'$ life time is about $1000\fm$ and the mean free path in dense
 hadronic matter exceeds $10\fm$ \cite{Kapusta:1996ww}, the final
 $\eta'$ multiplicity will exceed thermal model expectation.
 
 However, local chemical equilibrium between  $\eta'$
 and gluons may not be readily reached during the
 quark-gluon plasma evolution although quarks and gluons do
 reach local equilibrium very fast.  
 Therefore, $f_{\eta'}$ must evolve non-trivially in time even if the
 gluons are already locally equilibrated.
 To calculate such an effect, we assume that the gluon density is
 already thermal and rewrite the above as
 \be
 \partial_t f_{\eta'} + {\bf v}\cdot\nabla f_{\eta'}
 & = &
 {1\over 4E_P}
 \int {d^3 p\over (2\pi)^3 2 p} {d^3 q\over (2\pi)^3 2 q}\,
 \absol{M}_{gg\to\eta'}^2\, (2\pi)^4 \delta(p + q - P)
 \non
 & & {} \qquad \times
 f_g(p)\, f_g(q)\,
 \left[
 1 - f_{\eta'}(P)/f_{BE}(P) 
 \right]
 \ee
 where $f_{BE}(P) = 1/(e^{E_P/T} - 1)$.
 Substituting the matrix element yields
 \be
 \partial_t f_{\eta'} + {\bf v}\cdot\nabla f_{\eta'}
 & = &
 {\absol{H_0}^2 M^4_{\eta'}\over E_P}
 \left[
 1 - f_{\eta'}(P)/f_{BE}(P) 
 \right]\, \Gamma_2(P)
 \ee
 where the 2-body thermal phase space factor is given by
 \be
 \Gamma_2(P)
 =
 \int {d^3 p\over (2\pi)^3 2 p} {d^3 q\over (2\pi)^3 2 q}\,
 (2\pi)^4 \delta(p + q - P)
 f_g(p)\, f_g(q)\,
 \ee
 The evaluation of $\Gamma_2(P)$ can be found in
 Appendix~\ref{app:2body}.  In the Boltzmann limit,
 \be
 \Gamma_2(P) = {1\over 8\pi}\, e^{-E_P/T}
 \ee
 
 For simplicity, we take the Boltzmann limit from now on.
 As for the gluon evolution, we use hydrodynamic models with 1-D
 expansion to make a simple physical estimate. 
 In terms of 
 the space-time rapidity 
 $\eta = (1/2)\ln((t+z)/(t-z))$,
 the flow velocity in the 1-D Bjorken model is
 given by
 \be
 u^\mu = (\cosh\eta, 0, 0, \sinh\eta)
 \ee
 Using the ideal gas equation of motion $\epsilon = 3p$ results in a
 simple time dependence of the temperature
 \be
 T(\tau) = T_0\, \left(\tau_0 \over \tau\right)^{1/3}
 \label{eq:Ttau}
 \ee
 where $\tau = \sqrt{t^2 - z^2}$  is the proper time and
 $T_0$ is the temperature at the initial (proper) time $\tau_0$.
 Further, we limit the $\eta'$ momenta to be at the central rapidity so that 
 $v_z = P_z/E_P = 0$.
 In that case, only the time derivative term from 
 the left hand side of the Boltzmann equation remains non-vanishing.
 The coordinate $z$ and the momentum $P$ merely are parameters
 in 1-D ordinary differential equation
 \be
 {df(t, z, P)\over dt}
 =
 {\absol{H_0}^2 M^4_{\eta'}\over 8 \pi E_P}
 \left[
 e^{-E_P\cosh\eta/T(\tau)} - f(t, z, P) 
 \right]
 \ee
 where we have omitted the subscript label $\eta'$ from the
 distribution and $\eta$ and $T$ are functions of $t$ and $z$.
 This is in the form of a relaxation equation.
 The relaxation time is given by
 \be
 t_P^{\rm rel}
 =
 {8 \pi E_P \over \absol{H_0}^2 M^4_{\eta'}}
 =
 \tau_{\rm rel}\, \gamma_P 
 \ee
 where $\tau_{\rm rel} = 4.5\fm$ is the relaxation time in the rest
 frame of the $\eta'$ and
 $\gamma_P = E_P/M_{\eta'}$ is the Lorentz $\gamma$ factor
 associated with the $\eta'$ momentum.  
 Here we used $M_{\eta'} = 0.7\GeV$ in accordance with our earlier
 discussion. 
 Up to the momentum of
 $1\,\hbox{GeV}$, the typical $\gamma$ factor does not exceed 2.
 Therefore the relaxation time is comparable with
 the typical plasma life time of $1 - 10\,\hbox{fm}$. 
 The relaxation time $t_P^{\rm rel}$ is independent of the temperature unless
 $H_0$ and/or $M_{\eta'}$ depends strongly on $T$.  

 The solution of the above equation is given by
 \be
 f(t, z, P)
 =
 \int_{t_{\rm init}}^{t} {dt'\over t_P^{\rm rel}}\,
 e^{-(t-t')/t_P^{\rm rel}}\, f_0(t', z, P)
 \label{eq:f_sol}
 \ee
 with the initial condition $f(t_{\rm init}, z, P) = 0$ and 
 $f_0(t', z, P) = e^{-E_P\cosh\eta(t',z)/T(t',z)}$. 
 What we are interested in is the distribution function at the
 hadronization time $t_{\rm had}$.  
 In the Bjorken model, 
 the proper time at the hadronization is given by
 \be
 \tau_{\rm had} = \tau_0\, \left(T_0\over T_c\right)^3
 \ee
 We then take the initial proper time for the $\eta'$ evolution to be
 the larger of $\tau_0$ and
 \be
 \tau_{\rm init} = \tau_{\rm had} - \tau_{\rm rel}
 \ee
 The Mikowskian time and the proper time is related by
 \be
 t = \sqrt{\tau^2 + z^2}
 \label{eq:MinkwskianT}
 \ee
 Therefore, the farther away from the origin, the later the initial
 time is.  This is due to the strong longitudinal flow and time
 dilation associated with it.  
 The longitudinal flow is faster further away from the origin.
 It also means that at large $z$, there will be very little 
 time between the on-set of $\eta'$ production by fusing gluons and the
 hadronization time.

 \section{Numerical Results}

 To evaluate \eqn{eq:f_sol}, we take
 $T_0 = 0.334\GeV$ and $\tau_0 = 0.6\fm$.  These parameters are taken from
 a recent hydrodynamic study of the elliptic flow at RHIC\cite{Kolb:2001}.
 The initial temperature
 corresponds to the average energy density of about $23\GeV/\fm^{-3}$.
 The hadronization proper time with these parameters is 
 $\tau_{\rm had} = 4.6\fm$.
 Since 
 $\tau_{\rm had}-\tau_{\rm rel} = 0.1\fm$ is shorter than $\tau_0$,
 we set $\tau_{\rm init} = \tau_0$.

\begin{figure}[t]
  \epsfxsize=10cm
  \centerline{\epsfbox{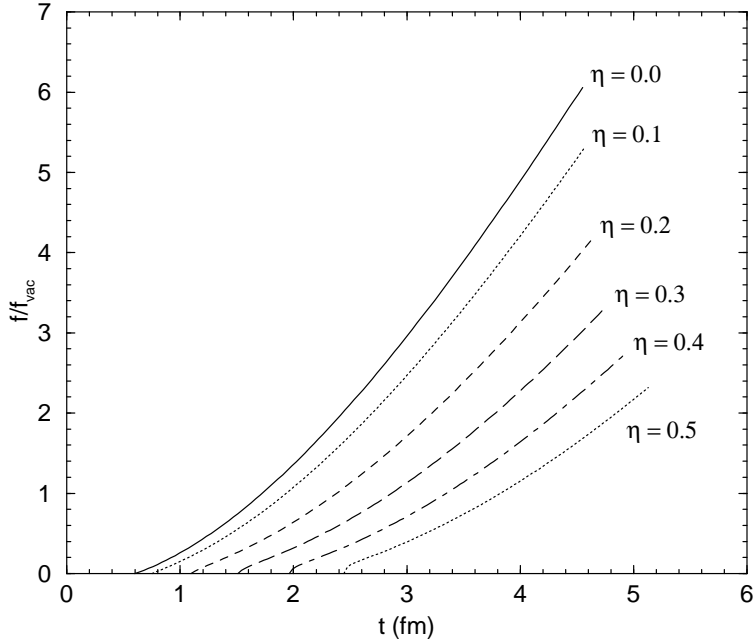}}
  \caption{
   The ratio of the solution of Eq.(\protect\ref{eq:f_sol}) 
   and $f_0 = e^{-M_0\cosh\eta(t,z)/T(t,z)}$ at $P = 0$ as a function of time.
   Here $M_0= 0.958\GeV$ is the vacuum $\eta'$ mass.
   From the bottom, the curves correspond to the 
   {\em final} space-time rapidities
   $\eta = 0.0$ through $0.5$ in steps of $0.1$ at $t_{\rm had}$.
   Calculated with the RHIC parameters.  The end points corresponds to
   $t_{\rm init} = \sqrt{\tau_{\rm init}^2 + z^2}$ and
   $t_{\rm fin} = \sqrt{\tau_{\rm fin}^2 + z^2}$.  
  }
  \label{fig:history}
 \end{figure}

 Fig.~\ref{fig:history} shows the numerical solutions with $P = 0$
 within the time interval $t_{\rm init} \le t \le t_{\rm had}$.
 In Fig.~\ref{fig:history} we plot the ratio of our solution and
 what one expects from the thermal model,
 \be
 f_{\rm vac} \equiv \exp\left(-M_0\cosh\eta(t,z)/T(t,z)\right)
 \label{eq:fvac}
 \ee
 where $M_0$ is the mass of $\eta'$ in vacuum.  

 The curves in Fig.~\ref{fig:history}
 starts from 0 and keeps growing.  This is due
 to two reasons.  One, the solution itself overshoots the equilibrium
 distribution $f_0$ (which has the same in-medium $M_{\eta'} = 0.7\GeV$)
 because the temperature is a decreasing function of time. 
 Initially the slope $df/dt$ is too steep for the eventual temperature
 of $T_c = 0.17\GeV$.
 Two, since the in-medium mass is 30\,\% smaller than the vacuum mass,
 $f_{\rm vac}$ in \eqn{eq:fvac} decreases much faster than either 
 $f$ or $f_0$ as $t$ increases.

 It is also apparent that for larger $\eta$ or equivalently
 larger $z$, there is not enough time
 between $t_{\rm init} = \sqrt{\tau_{\rm init}^2 + z^2}$ and 
 $t_{\rm fin} = \sqrt{\tau_{\rm fin}^2 + z^2}$ for the solution to grow
 over $f_{\rm vac}$.
 Consequently, the enhancement of 
 $dN_{\eta'}/dy$ at the mid-rapidity ($y=0$)
 \be
 {\left(dN_{\eta'}/dy\right)_{f} \over
 \left(dN_{\eta'}/dy\right)_{f_{\rm vac}}}
 \approx 2.5 
 \ \ \ \ \hbox{(RHIC)}
 \ee
 is not as large as the enhancement of $f$ at $\eta=0$.  
 Here we used
 \be
 \left. {dN_{\eta'}\over dy} \right|_{y=0}
 =
 \int {d^2P_T\over (2\pi)^3}\, E_P \int d^3x\, f(t_{\rm had}, z, P_T)
 \ee
 This enhancement factor is not particularly
 sensitive to the initial temperature.
 Keeping $\tau_0$ fixed, the ratio is 1.8 at $T_0 = 0.3\GeV$, increases
 up to 2.6 at $T_0 = 0.35\GeV$ and then
 decreases to 2.2 at $T_0 = 0.4\GeV$.
 
 One should note that this is {\em on top} of other processes 
 that produce $\eta'$ at the hadronization and in later times.
 Therefore, this result definitely indicates a large enhancement in the
 $\eta'$ yield due to the thermal gluon fusion process at RHIC.

 At LHC, the initial temperature can reach $T_0 = 1\GeV$.
 Accordingly, the hadronization takes place much later,
 $\tau_{\rm had} = 10 - 20\fm$ even though the equilibration 
 time is shorter, $\tau_0 \approx 0.1\fm$~\cite{Eskola:2000fc}.
 Therefore the rate of the change in the temperature is slower than the rate
 at RHIC (recall $T = T_0(\tau_0/\tau)^{1/3}$).  
 This implies that even though the initial temperature is much higher
 than the temperature at RHIC, the enhancement factor may not be much
 different.
 With $\tau_0$ fixed at $0.1\fm$, we get   
 \be
 {\left(dN_{\eta'}/dy\right)_{f} \over
 \left(dN_{\eta'}/dy\right)_{f_{\rm vac}}}
 \approx 2 - 3
 \ \ \ \ \hbox{(LHC)}
 \ee
 between $T_0 = 0.5\GeV$ and $T_0 = 1.0\GeV$. 
 The maximum enhancement factor 3 is reached at $T_0 = 0.6\GeV$.

 At SPS, the enhancement factor is more sensitive to the initial temperature.
 Keeping $\tau_0 = 0.8\fm$~\cite{Kolb:2001},
 the ratio increases as the temperature increases
 within $0.2\GeV \le T_0 \le 0.25\GeV$
 \be
 0.3 
 \lsim 
 {\left(dN_{\eta'}/dy\right)_{f} \over
 \left(dN_{\eta'}/dy\right)_{f_{\rm vac}}}
 \lsim 
 1.1
 \ \ \ \ \hbox{(SPS)}
 \ee
 Since $\eta'$ decays to $\eta\pi\pi$ in 65\,\% of the times, one may
 ask if this SPS result is compatible with the $\eta$ multiplicity 
 measurement by WA80 and the low mass dilepton spectrum measured by CERES.
 Thermal ratio of $\eta'$ and $\eta$ within the Bjorken scenario is
 17\,\%.  Therefore one would expect that about 11\,\% of $\eta$ comes
 from $\eta'$ decay.  Doubling that would indicate
 about 10\,\% increase in the $\eta$ multiplicity. 
 However, at present the experimental uncertainty is bigger
 than than 10\,\% \cite{Lebedev:1994rh,Albrecht:1995ug}. 

\begin{figure}[t]
  \epsfxsize=10cm
  \centerline{\epsfbox{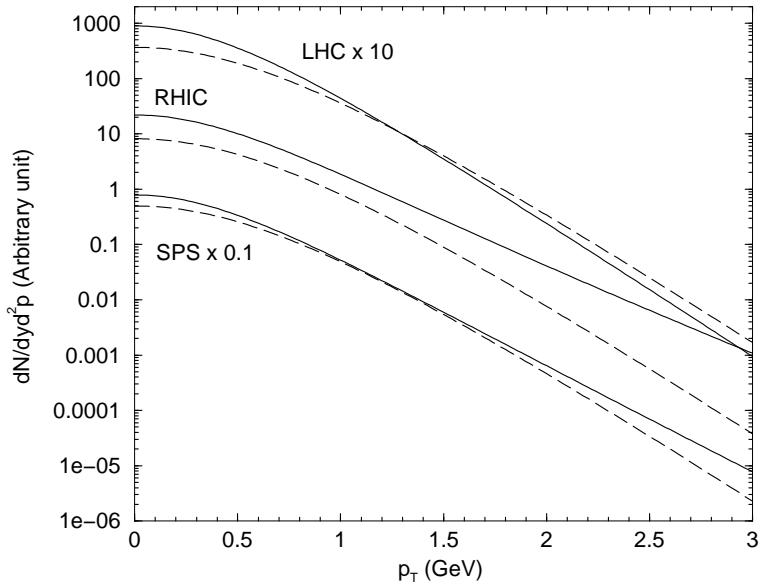}}
  \caption{
  The transverse momentum spectrum of $\eta'$ calculated with 
  the solution \eqn{eq:f_sol} (solid line) and 
  the thermal $e^{-E^0_P\cosh\eta/T_c}$ (broken line) where 
  $E^0_P$ is calculate with the vacuum $\eta'$ mass.
  For the top two lines,
  we used $T_0 = 1\GeV$ and $\tau_0 = 0.1\fm$ estimated for LHC in
  Ref.\cite{Eskola:2000fc}.
  SPS and RHIC parameters are $T_0 = 0.257\GeV, T_0 = 0.334\GeV$
  and $\tau_0 = 0.8\fm, \tau_0 = 0.6\fm$ respectively~\cite{Kolb:2001}.
  Transverse expansion is not taken into account.
  }
  \label{fig:dndydptpt}
 \end{figure}

 More detailed information than the yield can be obtained in the
 transverse momentum distribution shown in Fig.~\ref{fig:dndydptpt}.
 There is a clear difference between our calculation and
 the thermal distribution. 
 As one can see, the dependence of the solution \eqn{eq:f_sol} on 
 $T_0$ and $\tau_0$ is non-trivial.
 Since the simple 1-D model we employ does not take into account
 the transverse flow, the slope parameter
 of the $p_T$ spectra in Fig.~\ref{fig:dndydptpt}
 should be taken as qualitative estimates rather than quantitative
 predictions.

 \section{Discussion and Conclusion}

 In this paper, we calculated the yield and the momentum spectrum of the
 flavor singlet $\eta'$ mesons produced by the fusion of thermal gluons.
 It is shown above that at RHIC and LHC, there is a significant
 enhancement in $\eta'$ yield.  Furthermore, the $p_T$ spectrum of $\eta'$
 shows an interesting deviation from the $M_T$ scaling.
 The on-set of the deviation from the naive $M_T$ scaling contains 
 information on the initial conditions such as the initial
 temperature and the thermalization time.
 This may be feasible since $\eta'$ has a long life-time and
 a long mean free path.
 
 Further implication of our result includes the low mass dilepton
 enhancement.
 The branching ratio of $\eta' \to \eta \pi\pi$ is 65\,\%.
 A large number of $\eta'$ results in a sizable increase in 
 $\eta$ multiplicity which 
 in turn gives rise to the $\eta$ Dalitz peak in the dilepton 
 invariant mass spectrum.   At SPS, the enhancement is not significant
 enough to be noticed.  But at RHIC and LHC, there can be a substantial
 increase of the peak.
 Another observable where $\eta'$ enhancement plays a role is the HBT
 correlation.
 As shown in Ref.~\cite{Vance:1998wd}, enhanced $\eta'$ production
 reduces the strength of the HBT correlation at small $p_T$.  

 The above conclusions are based on the effective $gg\eta'$ 
 vertex deduced by Atwood and Soni, the Kinetic (Boltzmann) equation
 and also the following assumptions:
 \begin{itemize}
 \item[(i)] The gluons are locally thermalized and
 follow the hydrodynamic evolution.
 \item[(ii)] The strength of $gg\eta'$ vertex is independent of the
 temperature.
 \item[(iii)] The mass of $\eta'$ involved in this process 
 is lower than the vacuum value since only the $s$ quark loop is
 involved in the anomalous coupling.
 \item[(iv)] Kinetic equation is valid in this regime.
 \end{itemize}
 As to the validity of the hydrodynamics, 
 the measurements of the elliptic flow indicate the existence of
 collective motion.  
 Whether this implies thermal and chemical equilibrium is not entirely
 certain.   
 Since our result indicates that the $\eta'$ density is proportional to the
 square of the gluon density, the measured $\eta'$ $dN/dy$ must reflect the
 underlying gluon distribution be it thermal or the gluon $x$
 distribution function.
 For instance,
 if the plasma is gluon dominated right up to the hadronization,
 then one should see even more $\eta'$ than what we have estimated.

 One may also question the validity of the 1-D expansion model we employed.
 The full 3-D calculation with a realistic equation of state is 
 clearly out of the scope of this paper.  
 However, our main results should be robust 
 since faster falling temperature makes the $\eta'$
 distribution overshoot even more. 

 The temperature dependence of the $gg\eta'$ vertex and properties of
 $\eta'$ itself are at present not fully understood.
 In this study, we made simple assumptions to make
 a progress.  
 As we argued in Introduction, the strength of $gg\eta'$ vertex
 will not vanish as the strength of $\pi^0\gamma\gamma$ vertex does at
 $T_c$.  If one accepts the rough estimate \eqn{eq:rough_est} at
 the face value, then the strength of the $gg\eta'$ coupling could
 even be larger than $H_0$.
 The exact temperature dependence of the vertex, however, has yet to be
 worked out.  
 The crucial question is then:
 What exactly is the relation between $H_0$ and $M_{\eta'}$ as a function of
 temperature?
 High statistics measurement of $\eta'$ spectrum can potentially answer
 this question by employing the ideas developed in this paper.

 The kinetic equation description is valid when the mean free path
 is much larger than any other length scale.  This is certainly the case.  
 The mean
 free time of the $\eta'$ is longer than $4\fm$.  The dense medium at 
 $T = 300\MeV$ has much smaller inter-particle distance.
 The effect of inclusion of other processes such as $q\bar q\to \eta'$
 can be roughly estimated by raising the value of $H_0$.  
 In our calculation, this leads to more $\eta'$
 production by shortening $t_{\rm rel}$.
 
 In summary, we have shown that $\eta'$ is a good probe of 
 the gluons density using the recently proposed $gg\eta'$
 effective vertex.  Other application along the same idea includes the
 investigation of the in-medium properties of $\eta'$ and its possible link
 to the fate of the axial anomaly in a quark-gluon plasma.  
 It will be also interesting to study formation of $\eta'$ within gluon
 jets.  
 These and other aspects are currently under investigation.

 \section*{Acknowledgment}

 The author is grateful to S.~Pratt, J.~Jalilian-Marian, L.~McLerran, 
 A.~Soni, C.~Gale, D.~Kharzeev and S.~Bass 
 or helpful suggestions and discussions.
 This work was supported in part by the Natural Sciences and
 Engineering Council of Canada and by le Fonds pour la Formation
 de Chercheurs et l'Aide \`a la Recherche du Qu\`ebec.

 \appendix

 \section{Real Part of the $\eta'$ Self-Energy in Thermal Gluons}
 \label{app:self_energy}

 The Feynman diagrams for the one-loop retarded self-energy 
 of the $\eta'$ in equilibrium is given in Fig.~\ref{fig:self_energy}. 
 These can be calculated in many ways.  In this paper,
 we adopt the set of Feynman rules derived in 
 Ref.~\cite{Jeon:1998zj}.
 The diagrams then corresponds to the expressions
 \be
 \Sigma_{a}(P)
 =
 {-i\over 2}
 \int 
 {d^4 p\over (2\pi)^4}\,
 {d^4 q\over (2\pi)^4}\, (2\pi)^4\delta(p+q - P)\,
 (\Tr\,M(p,q)^2)\,
 G(p)\, G(q)
 \ee
 and 
 \be
 \Sigma_{b}(P)
 =
 {-i\over 2}
 \int 
 {d^4 p\over (2\pi)^4}\,
 {d^4 q\over (2\pi)^4}\, (2\pi)^4\delta(p+q - P)\,
 (\Tr\,M(p,q)^2)\,
  \Delta_+(p)\, \Delta_+(q) 
 \ee 
 where
 \be
 G(p)= {i\over p^2 + i\epsilon} + n_{BE}(p)2\pi\delta(p^2)
 \ee
 and
 \be
 \Delta_+(p)= \theta(p^0)2\pi\delta(p^2) + n_{BE}(p^0)2\pi\delta(p^2)
 \ee
 The prefactor $1/2$ comes from the fact that the two intermediate
 gluons are identical.
 Using these propagators, it is clear that the real part of the
 self-energy comes only from the diagram (a).  Hence we concentrate on
 evaluation of (a) from now on.
 
 First, we evaluate the vertex trace
 \be
 (\Tr\,M(p,q)^2)\,
 & = &
 \sum_{ab}\delta^{ab}
 \sum_{\lambda,\gamma} M^*_{\lambda\gamma} M^{\vphantom{*}}_{\lambda\gamma}
 \non
 & = & 
 8
 H_0^2\, 
 \epsilon_{\alpha\beta\mu\nu} \,
 \epsilon^{\alpha\beta\rho\sigma} \,
 p^\mu q^\nu p_\rho q_\sigma 
 \non
 & = &
 16H_0^2\,\left( (p\cdot q)^2 - p^2 q^2\right)
 \ee
 using the identities
 \be
 & \sum_\lambda (\epsilon_p^{\alpha})_\lambda^* (\epsilon_p^\zeta)_\lambda
 =
 -g^{\alpha\zeta} + a p^\alpha p^\zeta
 &
 \\
 \inline{and}
 &
 \epsilon_{\alpha\beta\mu\nu} \,
 \epsilon^{\alpha\beta\rho\sigma} 
 =
 2
 \left(g^\mu_\sigma\, g^\nu_\rho
 -
 g^\mu_\rho\, g^\nu_\sigma
 \right)
 &
 \ee
 The $a p^\alpha p^\zeta$ term does not contribute due to the
 anit-symmetric property of $\epsilon_{\alpha\beta\mu\nu}$.
 Then 
 \be
 \Sigma_{a}(P)
 & = &
 {-8i H_0^2}
 \int 
 {d^4 p\over (2\pi)^4}\,
 {d^4 q\over (2\pi)^4}\, (2\pi)^4\delta(p+q - P)\,
 \left( (p\cdot q)^2 - p^2 q^2\right)\,
 G(p)\, G(q)
 \non
 & = &
 {-8i H_0^2}
 \int 
 {d^4 p\over (2\pi)^4}\,
 \left( (p\cdot (P-p))^2 - p^2 (P-p)^2\right)\,
 G(p)\, G(P-p)
 \ee
 The zero-temperature part of this diagram is badly divergent.  In view
 of the effective theory nature of this vertex, we will simply drop the
 zero-temperature part
 in this calculation.  The thermal part can be separated into two
 parts,
 \be
 \Sigma_a(P) = \Sigma_{a,1}(P) + \Sigma_{a,2}(P) 
 \ee
 where 
 \be
 \Sigma_{a,1}(P)
 & = &
 {-16i H_0^2}
 \int 
 {d^4 p\over (2\pi)^4}\,
 (p\cdot P)^2 \,
 {i\over M^2 - 2P\cdot p + i\epsilon}\,
 n(p)2\pi\delta(p^2)
 \ee
 and
 \be
 \Sigma_{a,2}(P)
 & = &
 {-8i H_0^2}
 \int 
 {d^4 p\over (2\pi)^4}\,
 (p\cdot P)^2 \,
 n(E_{P-p})2\pi\delta(M^2 - 2P{\cdot} p)\,
 n(p)2\pi\delta(p^2)
 \ee
 and we used the on-shell conditions $p^2 = 0$ and $P^2 = M^2$.
 
 The real part of the self-energy comes only from 
 $\Sigma_{a,1}(P)$. 
 \be
 {\rm Re}\, \Sigma_{\rm ret}(P)
 & = &
 {\rm Re}\, \Sigma_{a,1}(P)
 \non
 & = &
 {16 H_0^2}
 \int 
 {d^4 p\over (2\pi)^4}\,
 (p\cdot P)^2 \,
 {\rm PP}\,
 {1\over M^2 - 2P\cdot p}\,
 n(p)2\pi\delta(p^2)
 \non
 & = &
 -{4 H_0^2 M^2}
 \int {d^4 p\over (2\pi)^4}\,
 n(p)2\pi\delta(p^2)
 \non
 & & {} 
 +
 {4 H_0^2 M^4}
 \int 
 {d^4 p\over (2\pi)^4}\,
 {\rm PP}\,
 {1\over M^2 - 2P\cdot p}\,
 n(p)2\pi\delta(p^2)
 \non
 \ee
 where ${\rm PP}$ signifies the principal part. 
 
 For simplicity, we orient $P^\mu = (E_P, 0, 0, P)$ and approximate
 the Bose-Einstein factor by a Boltzmann factor
 \be
 n(p) \approx e^{-p/T}
 \ee
 Then
 \be
 {\rm Re}\, \Sigma(P)
 & \approx &
 -{4 H_0^2 M^2}
 {1\over (2\pi)^2}
 \int dp\, p\, e^{-\absol{p}/T} 
 \non
 & & {} 
 +
 {4 H_0^2 M^4}
 \int 
 {d^4 p\over (2\pi)^4}\,
 {\rm PP}\,
 {1\over M^2 - 2P\cdot p}\,
 e^{-\absol{p}/T}\, 2\pi\delta(p^2)
 \non
 & = &
 -{4 H_0^2 M^2}
 {T^2\over (2\pi)^2}
 \Bigg[
 1 -
 \non
 & & {} 
 +
 {M^2\over 4 P T^2}
 \int 
 dp\,
 e^{-p/T}
 \left\{
 \ln
 \absol{
 M^2 - 2 E p + 2 P p
 \over
 M^2 - 2 E p - 2 P p
 }
 +
 \ln
 \absol{
 M^2 + 2 E p + 2 P p
 \over
 M^2 + 2 E p - 2 P p
 }
 \right\}
 \Bigg]
\non
 \ee
 The result can be expressed in terms of the exponential integral
 functions
 \be
 {\rm Re}\, \Sigma_{\rm ret}(P)
 & = &
 -M^2 {4 H_0^2 T^2\over (2\pi)^2}
 \left(1 - {M^2\over 4 PT} A(P)\right)
 \ee
 where
 \be
 A(P)
& = & 
{M^2}\,
e^{-(P+E_P)/2T}\,
\Bigg[ 
-e^{E_P/T}\,
        \hbox{Ei}\left({\frac{P - E_P}{2\,T}}\right)
         + e^{P/T}\,
      \hbox{Ei}\left({\frac{-P + E_P}{2\,T}}\right) 
\non & & \quad\qquad {}
+ 
     e^{(P + E_P)/T} \,
      \hbox{Ei}\left({\frac{-\left( P + E_P \right) }{2\,T}}\right) - 
     \hbox{Ei}\left({\frac{P + E_P}{2\,T}}\right)
      \Bigg] 
 \ee
 Numerically, between $T = 0.17\GeV$ and $T = 1.0\GeV$, the second term
 is important only near $P = 0$.  Therefore we approximate the above
 with
 \be
 {\rm Re}\, \Sigma_{\rm ret}(P)
 & \approx & \Sigma_{\rm ret}(P=0)
 \non
 & = &
 -M^2 {4 H_0^2 T^2\over (2\pi)^2}
 \left(1 
 - 
 {M^2\over 4 T} 
 \left(
 e^{M/2T} \hbox{Ei}(-M/2T)
 +
 e^{-M/2T} \hbox{Ei}(M/2T)
 \right)
 \right)
 \non
 \ee 
 We can self-consistently determine $M$ by solving the following
 equation for $M$ with $M_0 = 0.7\GeV$:
 \be
 M^2 = M_0^2 + {\rm Re}\,\Sigma_{\rm ret}(0)
 \ee
 Numerically solution of this equation indicates that $M(T)$ is a slow
 varying function of $T$.
 At $T = 0.17\GeV$, $M(T) = 0.7\GeV$ is indistinguishable from $M_0$. 
 Even at $T = 0.5\GeV$, $M(T) = 0.68\GeV$.  Only around 
 $T = 1\GeV$, $M(T) = 0.64\GeV$ is appreciably different from $M_0$.
 However, at this temperature, our Boltzmann approximation is no longer
 appropriate.  We can ignore the temperature dependence of the $\eta'$
 mass for our estimates.

 \section{2 Body Thermal Phase Space}
 \label{app:2body}

 We start from the expression
 \be
 \Gamma_2(P)
 =
 \int {d^3 p\over (2\pi)^3 2 p} {d^3 q\over (2\pi)^3 2 q}\,
 (2\pi)^4 \delta(p + q - P)
 f_g(p)\, f_g(q)\,
 \ee
 Carrying out the $d^3p$ integral and the $q$ angle integral yields
 \be
 \Gamma_2(P)
 & = &
 {1\over 4}
 \int
 {d^3 q\over (2\pi)^3 q \absol{{\bf P}-{\bf q}} }\,
 (2\pi) \delta(\absol{{\bf P}-{\bf q}} + q - E_P)
 f_g( \absol{{\bf P}-{\bf q}} )\, 
 f_g(q)
 \non
 & = &
 {1\over 4}{1\over 2\pi}
 {1\over \absol{{\bf P}}}\,
 \int_{q_{\rm min}}^{q_{\rm max}} 
 dq\,
 f_g( E_P - q )\, 
 f_g(q)
 \ee
 where
 \be
 & 
 \displaystyle
 q_{\rm min} = {M^2\over 2(E_P + P)}
 &
 \\
 \inline{and}
 &
 \displaystyle
 q_{\rm max} = {M^2\over 2(E_P - P)}
 &
 \ee
 Using the Bose-Einstein functions for $f$, we get
 \be
 \Gamma_2(P)
 & = &
 f_{BE}(E_P)\,
 {T\over 8\pi\absol{\bf P}}
 \bigg[
 \ln\left(e^{q_{\rm max}/T} - 1\right)
 -
 \ln\left(e^{q_{\rm min}/T} - 1\right)
 \non
 & & \qquad {}
 +
 \ln\left(e^{E_P/T} - e^{q_{\rm min}/T}\right)
 -
 \ln\left(e^{E_P/T} - e^{q_{\rm max}/T}\right)
 \bigg]
 \ee
 In $T\to 0$ limit, we recover the Boltzmann result
 \be
 \Gamma_2(P) = {1\over 8\pi} e^{-E_P/T}
 \ee

\end{document}